\documentclass{article}

\usepackage[letterpaper,margin=1in]{geometry}
\usepackage[numbers,sort&compress]{natbib}

\usepackage[utf8]{inputenc}
\usepackage[T1]{fontenc}
\usepackage{times}
\usepackage{hyperref}
\usepackage{url}
\usepackage{xurl}
\usepackage{booktabs}
\usepackage{tabularx}
\usepackage{array}
\usepackage{amsfonts}
\usepackage{amssymb}
\usepackage{microtype}
\usepackage{xcolor}
\usepackage{graphicx}
\usepackage{enumitem}
\usepackage{float}
\usepackage{caption}
\usepackage[section]{placeins}
\hypersetup{colorlinks=true,linkcolor=blue!50!black,citecolor=blue!50!black,urlcolor=blue!50!black}

\title{Buildrix: An Open Agentic AI Benchmark for\\Building Engineering}

\author{%
  \textbf{Development Team}\\
  Zixin Jiang\textsuperscript{1}, Zixuan Qi\textsuperscript{2}, Nan Ma\textsuperscript{3},
  Sicheng Zhan\textsuperscript{4},\\ Judah Goldfeder\textsuperscript{5}, Thomas Tang\textsuperscript{6},
  Han Li\textsuperscript{7}, Bing Dong\textsuperscript{8}\\[6pt]
  \textbf{Task Contributors}\\
  Bingtong Guo\textsuperscript{8}, You Jeong Kim\textsuperscript{9}, Hongjun Li\textsuperscript{10},
  Matthew P. So\textsuperscript{5}, Xiao Wang\textsuperscript{11}, Xuezheng Wang\textsuperscript{12}\\[6pt]
  \small\textsuperscript{1}Massachusetts Institute of Technology\quad
  \textsuperscript{2}Texas A\&M University\quad
  \textsuperscript{3}Worcester Polytechnic Institute\\
  \small\textsuperscript{4}The University of Hong Kong\quad
  \textsuperscript{5}Columbia University\quad
  \textsuperscript{6}NVIDIA\quad
  \textsuperscript{7}Lawrence Berkeley National Laboratory\\
  \small\textsuperscript{8}Syracuse University\quad
  \textsuperscript{9}Colorado School of Mines\quad
  \textsuperscript{10}The Pennsylvania State University\\
  \small\textsuperscript{11}Tsinghua University\quad
  \textsuperscript{12}Johnson Controls\\[4pt]
  \small Correspondence: \texttt{zjiang25@mit.edu}
}
\date{}

\begin{document}

\maketitle

\begin{abstract}
Agentic AI has demonstrated remarkable progress across science and engineering, including mathematics, physics, chemistry, and biology. However, how well AI agents can perform real-world building engineering tasks remains largely unexplored. To address this gap, we introduce \textbf{Buildrix}, an open benchmark for evaluating AI agents on real building engineering tasks across eight domains. We investigated three questions: \emph{how AI agents compare with human engineers, whether expert knowledge improves agent performance, and how engineers can collaborate effectively with AI agents.} The benchmark shows that AI agents achieved an average score of 78 out of 100, compared with 84 for human engineers, while completing tasks in an average of 9 minutes at a cost of \$0.58 per run, compared with 8--160 reported working hours for human engineers, demonstrating promising potential. Interestingly, providing expert knowledge did not consistently improve agent performance and sometimes led to worse results. In contrast, structured task instructions substantially improved performance, increasing the average score from 27 to 75 and the success rate from 5\% to 76\%. These findings highlight both the potential and limitations of AI agents in building engineering, as well as the importance of structured task specifications and clear deliverable requirements for effective human--agent collaboration. More broadly, \textbf{Buildrix} provides an open evaluation framework for tracking AI progress, identifying capability gaps, and guiding the development of more capable and reliable AI agents to accelerate the adoption of AI in building engineering practice.
\end{abstract}

\section{Introduction}
\label{sec:intro}

With rapid urbanization, modern buildings are becoming increasingly complex and interconnected ecosystems that play a critical role in human well-being, energy use, and climate change. People spend nearly 90\% of their time in buildings, where they live, study, work, and interact with others~\cite{klepeis2001nhaps}. Buildings must therefore provide healthy and comfortable environments. They also account for about 30\% of global final energy use and a large share of energy-related CO\textsubscript{2} emissions~\cite{iea2022buildings}, making them critical to achieving net-zero emissions by 2050. At the same time, climate change and increasingly frequent extreme weather events place growing demands on building safety and resilience.

Meeting these demands requires substantial engineering effort throughout the building lifecycle, from design and construction to commissioning, operation, maintenance, and retrofit. Much of this work is time-consuming, labor-intensive, and heavily dependent on expert knowledge and practical experience~\cite{coakley2014calibration,zhang2025agenticbem}, which limits the progress of building renovation at a large scale.

In recent years, AI agents have demonstrated strong capabilities across science and engineering. In mathematics, they reached gold-medal performance at the 2025 International Mathematical Olympiad~\cite{deepmind2025imo}. In physics, they have helped derive new theoretical results~\cite{guevara2026gluon}. In chemistry and materials science, agents have planned experiments and supported the discovery and synthesis of new compounds~\cite{boiko2023autonomous,veiga2026materials}. In biology, they have proposed drug repurposing candidates and designed nanobodies with laboratory validation~\cite{gottweis2025coscientist,swanson2025virtuallab}. In software engineering, evaluations in early 2026 reported about 50\% success on tasks that take skilled engineers about 12 hours~\cite{kwa2025longtasks,metr2026frontier}. More broadly, Humanity's Last Exam evaluates models on 2,500 expert-level questions across more than 100 subjects. Reported scores for leading models rose from below 10\% in early 2025 to above 60\% by November 2026~\cite{phan2025hle,aa2026hle}.

These advances raise a natural question: How well can AI agents perform real-world building engineering tasks? Building engineering involves multi-scale physical systems, long-term iterative workflows, noisy sensor data, complex geometric drawings, specialized software, and professional judgment. These characteristics make it a particularly challenging domain for AI agents. Success in other scientific and engineering benchmarks does not necessarily translate into the ability to solve practical building engineering problems. Understanding their capabilities and limitations requires systematic evaluation using real engineering tasks, human reference solutions, and consistent assessment criteria.

To address this gap, we introduce Buildrix, an open benchmark for evaluating AI agents on real building engineering tasks, spanning eight domains across the building lifecycle: performance modeling and simulation; design, retrofit, and decarbonization; operations, control, and optimization; fault detection and commissioning; occupants, comfort, and indoor environment; energy forecasting and analytics; grid-interactive and integrated energy; and data, semantics, and digital twins~\cite{ashrae_tcs,iea_ebc_projects} (Figure~\ref{fig:3}).

Through a systematic collection and evaluation of tasks drawn from real engineering projects, Buildrix aims to answer three questions:

\begin{itemize}[leftmargin=*,itemsep=2pt,topsep=0pt,parsep=0pt]
\item \textbf{Agent vs. human.} How do AI agents compare with engineers on real building engineering tasks in terms of quality, time and cost?
\item \textbf{Human to agent.} Does expert knowledge improve agent performance? When does it help, and which types of knowledge matter most?
\item \textbf{Human with agent.} How can engineers collaborate with AI agents to improve their efficiency?
\end{itemize}

\begin{figure}[htbp]
  \centering
  \includegraphics[width=0.6\linewidth]{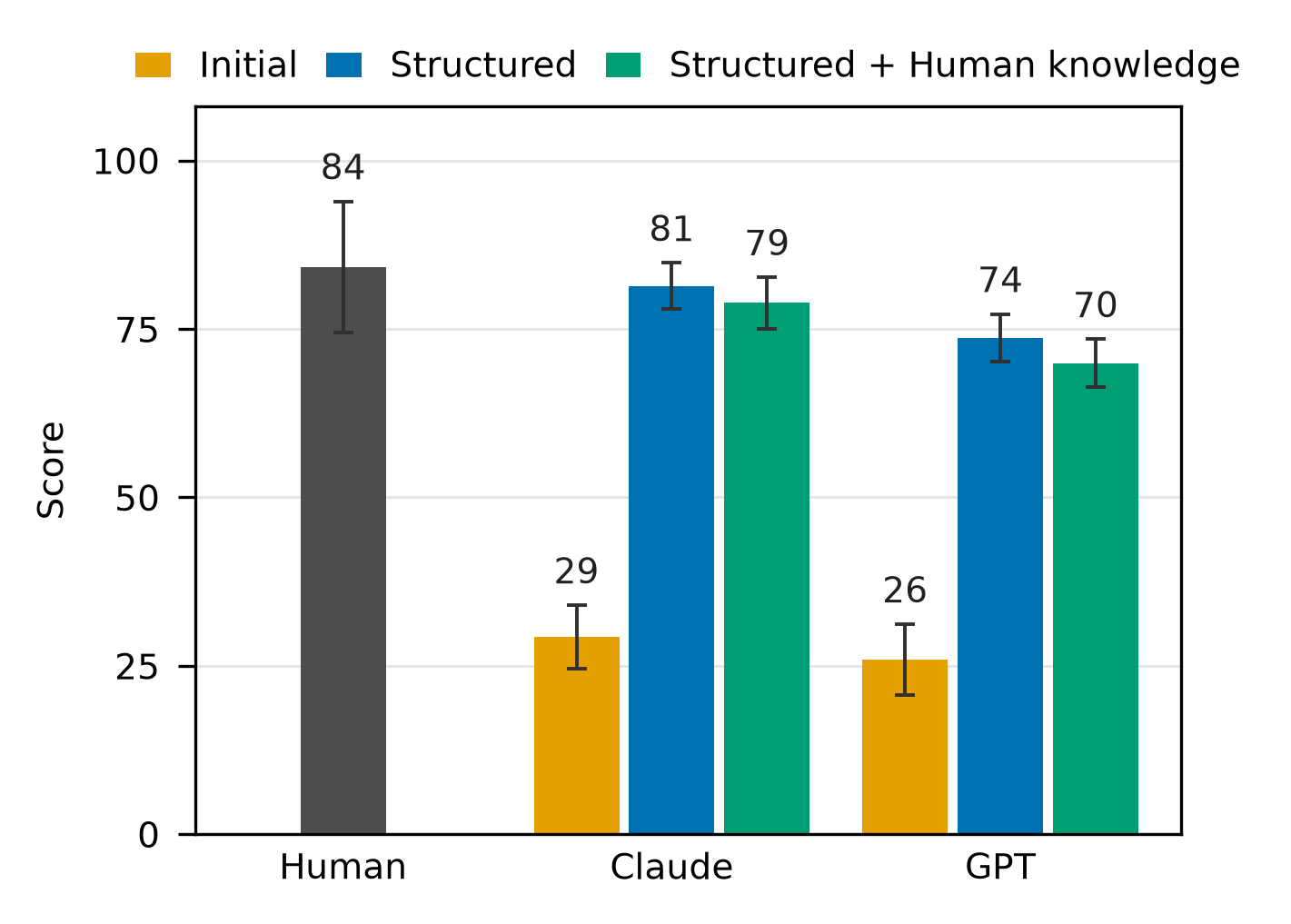}
  \caption{Mean scores of human engineers and Claude and GPT agents across different models and reasoning levels under three versions of each task: the contributor's initial request, the LLM-assisted structured request, and the structured request supplemented with expert knowledge. Error bars represent the standard error of the mean.}
  \label{fig:1}
\end{figure}

The initial benchmark evaluation (Figure~\ref{fig:1}) reveals both promising capabilities and important limitations. AI agents achieved an average score of 78 out of 100, compared with 84 for human engineers, while completing tasks in an average of 9 minutes at a cost of \$0.58 per run. However, providing expert knowledge did not consistently improve agent performance. In contrast, structured task requests substantially improved performance, increasing the mean score from 27 to 75 and the success rate from 5\% to 76\%.

Beyond these initial findings, Buildrix provides a reusable and extensible evaluation framework for the research community to assess AI capabilities, track progress, identify limitations, and guide the development of more capable and reliable AI agents for building engineering. Figure~\ref{fig:2} gives an overview of the benchmark.

The rest of the paper is organized as follows. Section 2 reviews related work and outlines the contributions of this study. Section 3 describes the benchmark and Section 4 the evaluation protocol. Section 5 presents the results, Section 6 discusses them, and Section 7 concludes.

\section{Related Work and Contributions}
\label{sec:related}

\subsection{Agent benchmarks in building engineering}

Research on large language models (LLMs) in building engineering spans energy modeling, HVAC control, fault detection, and building information modeling (BIM)~\cite{liu2025bldgllm,wan2025genai,ma2026tenq,neubauer2026hvac}. Most evaluations in this area test individual applications: answering technical questions about building systems and energy simulation~\cite{liang2025aecbench}, extracting energy model inputs from documents~\cite{jia2026bemeval}, querying BIM data~\cite{lin2026qwenbim}, and editing BIM models~\cite{nithyanantham2026bimedit}. More recent studies evaluate agents that use tools to carry out building and energy workflows (Table~\ref{tab:1}). Dieng et~al. evaluate agents that turn HVAC instructions from four user personas into EnergyPlus measures~\cite{dieng2026hvacsim}; EnergyBridge evaluates household demand-response planning with simulated residents~\cite{wu2026energybridge}; and Akinpelu et~al. evaluate energy-market analytics with live data, scored by LLM judges~\cite{akinpelu2026energy}. These studies establish evaluation methods for specific applications, \textbf{but do not establish how agents compare with engineers in output quality, time, and cost on the same tasks}. A review of 66 studies on LLMs for HVAC operations likewise calls for benchmarks validated in real-world settings~\cite{neubauer2026hvac}.

\begin{table}[htbp]
  \centering
  \scriptsize
  \caption{Agent benchmarks in building engineering.}
  \label{tab:1}
  \setlength{\tabcolsep}{3pt}
  \begin{tabularx}{\linewidth}{>{\raggedright\arraybackslash\hsize=0.85\hsize}X>{\raggedright\arraybackslash\hsize=1.05\hsize}X>{\raggedright\arraybackslash\hsize=0.75\hsize}X>{\raggedright\arraybackslash\hsize=0.9\hsize}X>{\raggedright\arraybackslash\hsize=0.9\hsize}X>{\raggedright\arraybackslash\hsize=0.9\hsize}X>{\raggedright\arraybackslash\hsize=1.65\hsize}X}
  \toprule
  \textbf{Benchmark} & \textbf{Tasks} & \textbf{Scale} & \textbf{Task source} & \textbf{Metric} & \textbf{Human baseline} & \textbf{Task specification and guidance} \\
  \midrule
  Dieng et~al.~\cite{dieng2026hvacsim} & Turn HVAC instructions into EnergyPlus measures & 1,320 instructions & -- & Translation accuracy & -- & Instructions representing four user types \\
  EnergyBridge~\cite{wu2026energybridge} & Plan household demand response with simulated residents & 350 episodes & Simulated homes and users & Acceptance, delivered flexibility & -- & Simulated user approval \\
  Akinpelu et~al.~\cite{akinpelu2026energy} & Answer energy-market analytics questions with live data & 243 problems & Industry practitioners & LLM judges & -- & -- \\
  \midrule
  \textbf{Buildrix (ours)} & Deliver complete engineering projects in building domains & 8 pilot projects, growing & Practitioners' completed projects & Task rubric score (0--100), success rate & Engineer's result, hours, cost & Initial vs. structured requests; added expert knowledge \\
  \bottomrule
  \end{tabularx}
\end{table}

\subsection{Agent benchmarks in AI}

Agent benchmarks in AI cover software engineering and computer use~\cite{jimenez2024swebench,zhou2024webarena,mialon2024gaia,xie2024osworld,liu2024agentbench}, machine learning and scientific research~\cite{chan2025mlebench,wijk2024rebench,starace2025paperbench,chen2025scienceagentbench}, professional work~\cite{patwardhan2025gdpval,mazeika2025rli,xu2024theagentcompany,sun2026ale}, and long-horizon tasks~\cite{rein2025hcast,kwa2025longtasks}. Together, they provide a solid foundation for measuring task completion, output quality, and performance relative to human experts.

A second line of work studies how the information given to agents affects their performance. GDPval compares full and shortened task prompts~\cite{patwardhan2025gdpval}, and Ambig-SWE removes key task details that agents can recover only by asking a simulated user~\cite{vijayvargiya2026ambig}. Both create incomplete requests by reducing a fuller specification, so they do not directly measure the benefit of clarifying a practitioner's initial request. Related benchmarks evaluate clarification-seeking and human--agent collaboration~\cite{edwards2026askorassume,li2026clareval,wang2025learningtoask,yao2024taubench,shao2025collabgym,shao2026collabskill}; CollabSkill, for example, pairs real workers with agents~\cite{shao2026collabskill}. Other studies vary the knowledge available to agents: ScienceAgentBench compares performance with and without expert knowledge~\cite{chen2025scienceagentbench}, and SkillsBench evaluates knowledge packaged as reusable agent skills~\cite{li2026skillsbench}. Curated skills raised the SkillsBench pass rate from 33.9\% to 50.5\%~\cite{li2026skillsbench}, but 92 of the 98 skills tested in SWE-Skills-Bench had no statistically significant effect~\cite{han2026sweskills}. It therefore remains unclear whether an engineer's guidance improves agent performance once the task requirements are made clear.

Table~\ref{tab:2} compares representative benchmarks by task source, metric, human baseline, and how each varies the task specification or guidance given to the agent.

\begin{table}[htbp]
  \centering
  \scriptsize
  \caption{Agent benchmarks.}
  \label{tab:2}
  \setlength{\tabcolsep}{3pt}
  \begin{tabularx}{\linewidth}{>{\raggedright\arraybackslash\hsize=0.85\hsize}X>{\raggedright\arraybackslash\hsize=1.05\hsize}X>{\raggedright\arraybackslash\hsize=0.75\hsize}X>{\raggedright\arraybackslash\hsize=0.9\hsize}X>{\raggedright\arraybackslash\hsize=0.9\hsize}X>{\raggedright\arraybackslash\hsize=0.9\hsize}X>{\raggedright\arraybackslash\hsize=1.65\hsize}X}
  \toprule
  \textbf{Benchmark} & \textbf{Domain} & \textbf{Scale} & \textbf{Task source} & \textbf{Metric} & \textbf{Human baseline} & \textbf{Task specification and guidance} \\
  \midrule
  SWE-bench~\cite{jimenez2024swebench} & Fixing Python repository issues & 2,294 issues, 12 repos & Real GitHub issues & \% resolved by tests & -- & -- \\
  WebArena~\cite{zhou2024webarena} & Web navigation on self-hosted websites & 812 tasks & Written by authors & Success rate & Graduate students, 78\% success & -- \\
  GAIA~\cite{mialon2024gaia} & General-assistant questions requiring tools & 466 questions & Human annotators & Exact-answer accuracy & Human respondents, 92\% accuracy & -- \\
  MLE-bench~\cite{chan2025mlebench} & ML engineering competitions & 75 competitions & Kaggle & Medal rate & Human Kaggle entrants & -- \\
  RE-Bench~\cite{wijk2024rebench} & ML research engineering & 7 tasks & Written by authors & Normalized score & 61 experts, 8 h each & -- \\
  PaperBench~\cite{starace2025paperbench} & Replicating ML papers & 20 papers & ICML papers & Replication score, LLM judge & ML PhDs, 48 h & -- \\
  HCAST~\cite{rein2025hcast} & Software, ML, security tasks & 189 tasks & Staff and contractors & Success vs. human time & 563 timed human attempts & -- \\
  GDPval~\cite{patwardhan2025gdpval} & Work products, 44 occupations & 1,320 tasks & Industry experts & Expert win rate & Expert deliverables, time, cost & Full vs. shortened task prompts \\
  Remote Labor Index~\cite{mazeika2025rli} & Freelance projects & 240 projects & Real freelance jobs & Automation rate & Freelancer deliverables & -- \\
  Agents' Last Exam~\cite{sun2026ale} & Professional workflows & 1K+ tasks, 55 fields & Industry experts & Pass rate, score & -- & -- \\
  Science\-Agent\-Bench~\cite{chen2025scienceagentbench} & Data-driven science coding & 102 tasks & Published papers & Success rate & -- & With vs. without expert knowledge \\
  Ambig-SWE~\cite{vijayvargiya2026ambig} & Underspecified software issues & 500 issues & SWE-bench Verified & \% resolved & -- & Key details removed; agent may ask a simulated user \\
  $\tau$-bench~\cite{yao2024taubench} & Customer-service dialogues & 165 tasks & Written by authors & Pass\textsuperscript{$k$} (consistency over k trials) & -- & Requirements revealed by a simulated user in dialogue \\
  CollabSkill~\cite{shao2026collabskill} & Occupational tasks done with human partners & 165 tasks, 93 workers & GDPval, APEX & Skill rating, rubric & Real workers with agents & Agent alone vs. agent working with a person \\
  SkillsBench~\cite{li2026skillsbench} & Tasks in 8 domains & 87 tasks & Community contributors & Pass rate & -- & No skills vs. curated vs. self-generated skills \\
  EngDesign~\cite{guo2025engdesign} & Engineering design, 9 domains & 101 tasks & Graduate students, researchers & Simulation-verified pass rate & -- & -- \\
  \midrule
  \textbf{Buildrix (ours)} & Complete building engineering projects & 8 pilot projects, growing & Practitioners' completed projects & Task rubric score (0--100), success rate & Engineer's result, hours, cost & Initial vs. structured request; with vs. without expert knowledge \\
  \bottomrule
  \end{tabularx}
\end{table}

Taken together, existing work leaves two gaps.

\begin{itemize}[leftmargin=*,itemsep=2pt,topsep=0pt,parsep=0pt]
\item First, to our knowledge, no benchmark evaluates agents on complete, end-to-end projects in building engineering, where tasks combine physical modeling, noisy measured data, specialized software, and professional judgment.
\item Second, benchmarks in other fields study human baselines, request clarity, and expert knowledge separately; to our knowledge, none examines all three on the same real tasks.
\end{itemize}

As a result, three practical questions remain open: how agents compare with engineers in quality, time, and cost; whether expert knowledge improves agent performance; and how the way a task is specified affects what agents deliver. Buildrix addresses both gaps by keeping, for each real project, the engineer's own result, the initial request, a structured request, and the engineer's added knowledge.

\subsection{Our Contributions}

\begin{itemize}[leftmargin=*,itemsep=2pt,topsep=0pt,parsep=0pt]
\item We introduce Buildrix, the first open benchmark for evaluating AI agents on real building engineering projects across the building lifecycle, providing a common basis for measuring and advancing AI capabilities in this domain.
\item We compare current AI agents directly with human engineers on quality, success rate, time, and cost, quantifying what agents can and cannot yet do in building engineering and providing evidence to guide AI development and practical adoption.
\item We examine how task specification and expert knowledge affect agent performance, identifying barriers to, and opportunities for, effective human--agent collaboration and delegation of engineering work.
\item We release the tasks, evaluation methods, and reusable engineering knowledge as an open, growing resource, so that the community can identify capability gaps, develop better agents, and track progress over time.
\end{itemize}

\section{Benchmark Design and Task Collection}
\label{sec:benchmark}

\begin{figure}[htbp]
  \centering
  \includegraphics[width=\linewidth]{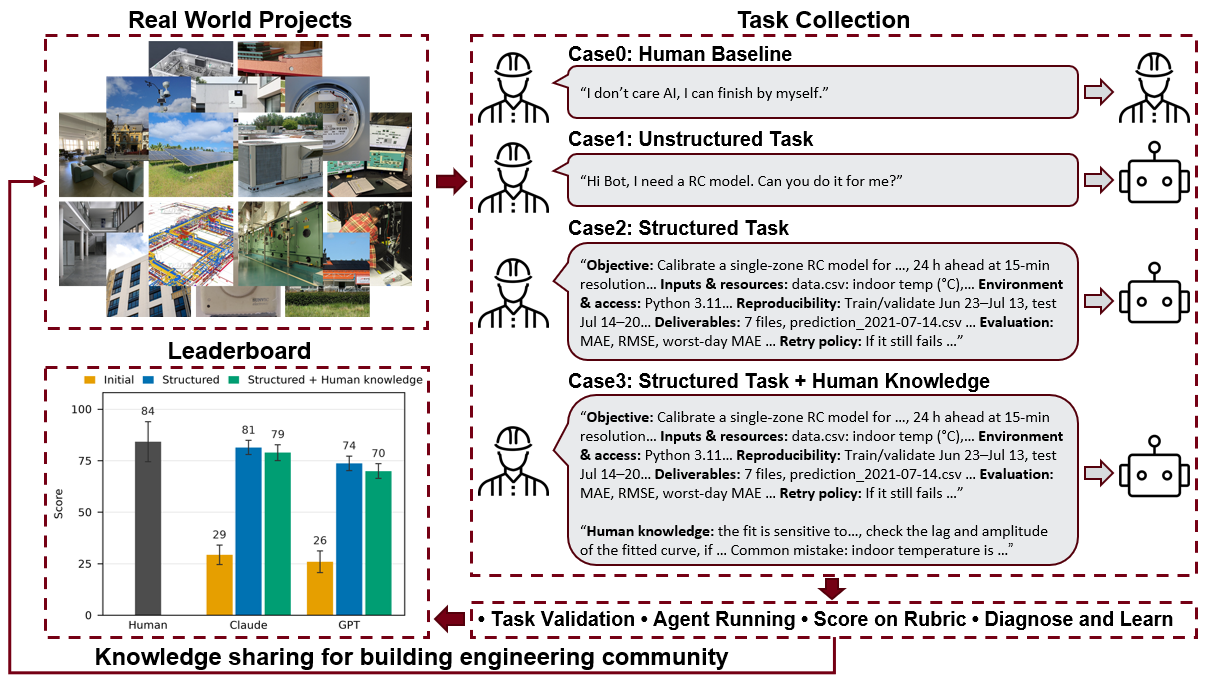}
  \caption{Overview of Buildrix. Tasks derived from real world projects that engineers have already completed. Each task can be done by the engineer (Case 0, the human baseline) or assigned to an agent in three ways: the initial request (Case 1), a structured request clarified from seven dimensions (Case 2), and the structured request with the engineer's knowledge (Case 3). Every agent run is scored with a rubric validated on the engineer's own result and diagnosed, and the results go to a public leaderboard and to lessons shared with the building engineering community.}
  \label{fig:2}
\end{figure}

\subsection{Overview}

Buildrix is designed to evaluate AI agents on real-world building engineering tasks across the building lifecycle, which collects tasks derived from engineering projects completed by human experts. Each task requires agents to produce practical engineering deliverables, such as calibrated models, energy forecasts, or optimized control schedules.

To ensure that tasks are suitable for systematic evaluation, each accepted task must meet four requirements:

\begin{itemize}[leftmargin=*,itemsep=2pt,topsep=0pt,parsep=0pt]
\item \textbf{Authenticity.} The task originates from a real engineering project rather than being created specifically for the benchmark.
\item \textbf{Completeness.} The task provides sufficient information, input data, and resources for an engineer or AI agent to complete the assigned work.
\item \textbf{Verifiability.} The task has clearly defined evaluation criteria and a quantitative scoring rubric that converts the quality of engineering deliverables into a score from 0 to 100.
\item \textbf{Human validation.} The task has been completed by a human practitioner, whose results provide a reference for verifying task feasibility and evaluation criteria.
\end{itemize}

To investigate how task specifications and human knowledge affect agent performance, Buildrix preserves three versions of each task: (1) the contributor's initial request (which in most cases is unclear or ambiguous), (2) an LLM-assisted structured request that clarifies task requirements along seven dimensions, and (3) the structured request supplemented with expert knowledge. These three versions allow us to evaluate agent performance under different levels of task specification and human guidance, and the contributor's own result serves as the human baseline (Figure~\ref{fig:2}).

The following sections describe the task domains and the procedure used to collect, structure, and validate engineering tasks.

\subsection{Task domains}

\begin{figure}[htbp]
  \centering
  \includegraphics[width=0.55\linewidth]{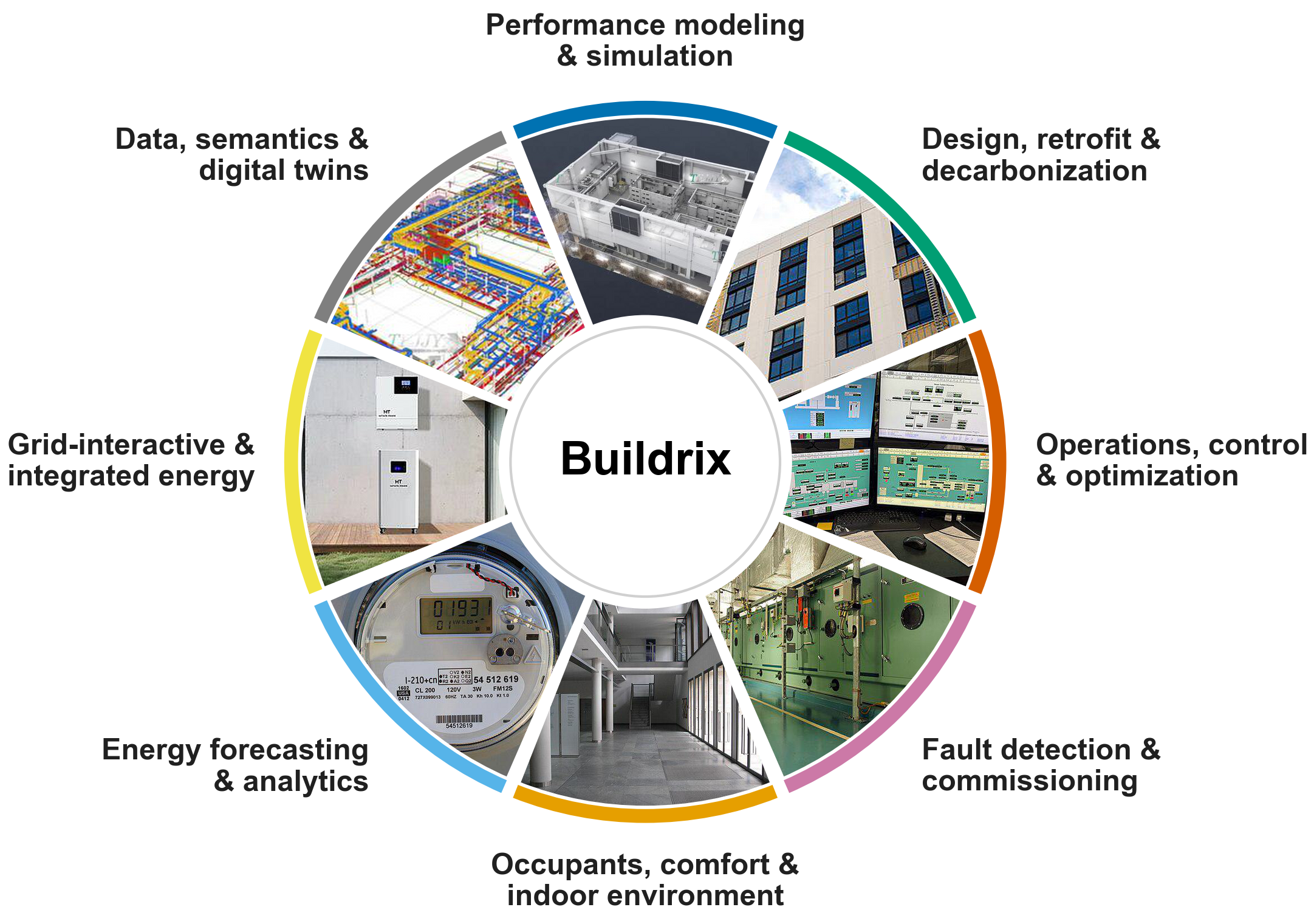}
  \caption{The eight Buildrix task domains, defined from the ASHRAE technical committees~\cite{ashrae_tcs} and the research projects of the IEA Energy in Buildings and Communities Programme~\cite{iea_ebc_projects}.}
  \label{fig:3}
\end{figure}

Building engineering involves a wide range of activities across different stages of the building lifecycle, from initial design and performance modeling to operation, maintenance, and retrofit. To represent this diversity, Buildrix organizes engineering tasks into eight domains (Figure~\ref{fig:3}), defined from the ASHRAE technical committees~\cite{ashrae_tcs} and the research projects of the IEA Energy in Buildings and Communities Programme~\cite{iea_ebc_projects}.

\subsection{Task construction}

Buildrix collects tasks from practitioners who have already completed the corresponding work in real engineering projects. Each task is processed through three steps to preserve the original engineering request, establish a clear and reproducible task specification, and collect additional human knowledge.

\textbf{Step 1: Collect the initial request.} Contributors describe the task as clearly and fully as possible, as if assigning it to a colleague, and provide the necessary input files and resources. The original request is preserved before any clarification or restructuring. Contributors also provide their completed work as a human reference for subsequent task validation and evaluation.

\textbf{Step 2: Clarify and structure the task.} An LLM reviewer examines the initial submission and checks whether seven dimensions are clearly specified:

\begin{itemize}[leftmargin=*,itemsep=2pt,topsep=0pt,parsep=0pt]
\item \textbf{Objective.} The goal and scope clearly explain what the agent is expected to accomplish.
\item \textbf{Inputs and resources.} All required data, files, and supporting materials are available, with descriptions of their contents, units, and intended use.
\item \textbf{Environment and access.} Required software, versions, licenses, computing resources, and access conditions are specified.
\item \textbf{Reproducibility.} Task configurations, including data periods, splits, settings, and random seeds where applicable, are specified to support reproducible and comparable results.
\item \textbf{Deliverables.} The required output files, formats, and expected contents are clearly defined.
\item \textbf{Evaluation.} A quantitative scoring rubric defines how the engineering deliverables are evaluated on a scale from 0 to 100, including the metrics, calculation methods, weights, and passing thresholds.
\item \textbf{Retry policy.} Rules specify when the agent may retry after failure, the allowed number of attempts or candidate approaches, stopping conditions, and required final outputs.
\end{itemize}

The reviewer classifies each dimension as clear, vague, or missing and asks follow-up questions to resolve unclear or missing information. The contributor responds, and the reviewer updates the description. This process continues until all seven dimensions are clearly specified. The reviewer then consolidates the agreed requirements into a standardized, self-contained task description. The contributor's reference results are maintained separately for evaluation rather than provided as solutions to the agent.

\textbf{Step 3: Collect human knowledge.} After the structured task description is finalized, contributors provide additional knowledge that they believe could help an AI agent complete the task. This information is flexible and may include recommended procedures, practical tips, intermediate checkpoints, common mistakes and fixes, useful software packages, scripts, or reference code. Unlike the essential task requirements collected in Step 2, this information represents optional expert guidance on how to approach and solve the problem.

The three finalized task versions include the initial request, the structured request, and the structured request supplemented with human knowledge. Together, they provide a basis for evaluating how task specifications and expert guidance affect AI agent performance.

\section{Evaluation Protocol}
\label{sec:protocol}

Buildrix follows a standardized evaluation protocol to ensure consistent and reproducible comparisons across AI agents and human engineers. Each agent receives the same task-specific inputs, operates under common time and resource constraints, and is evaluated using a quantitative scoring script validated against the human engineer's reference results. The protocol consists of three parts: agent execution, validated scoring, and failure diagnosis (Figure~\ref{fig:2}, bottom).

\subsection{Agents and runs}

We evaluated six AI models: Claude Haiku 5.5, Sonnet 5.5, and Opus 5.5 from Anthropic, and GPT-5.6 Luna, Terra, and Sol from OpenAI. Each model was evaluated at three reasoning levels (low, medium, and high). Claude models were executed using Claude Code, while GPT models used the Codex CLI.

To ensure consistent evaluation, all tasks were finalized and stored in a fixed study bundle, with checksums recorded for each task package before execution. Each run was conducted in a fresh, isolated workspace containing only the inputs required for the assigned task version. Other tasks, previous runs, ground-truth data, and human reference results were kept outside the workspace. Agents started with clean home directories and configurations, without personal settings, memory, or skills. All agents had access to the same pre-installed tools and operated without human intervention.

\subsection{Validated scoring}

Buildrix uses task-specific quantitative scoring scripts to evaluate engineering deliverables. Before agent evaluation, an AI grader generates a scoring script based on each task's predefined evaluation rubric. The script is first applied to the human contributor's reference results to verify that the calculated metrics match the reported values, the required ground-truth data are available, and the reference deliverables satisfy all passing thresholds. Tasks are accepted only after these checks are passed. The validated scoring script is then applied to every agent run, evaluating its generated deliverables against the ground-truth data. Each run receives a numerical score from 0 to 100 and a binary success indicator. A run is considered successful only if all required deliverables are provided and all passing thresholds are satisfied.

\subsection{Failure diagnosis}

Beyond quantitative scoring, Buildrix uses an LLM grader to analyze unsuccessful runs and identify common challenges faced by agents. The grader examines each run's evaluation results, generated outputs, and execution transcript, then provides a brief diagnosis and assigns one primary failure category: the task (unclear requirements), the environment (tool or infrastructure failure) or the agent (execution mistake, weak method or unfinished run). For each task, the grader also summarizes common failure patterns, approaches used by successful runs, and practical lessons for improving agent performance.

\section{Results}
\label{sec:results}

\subsection{Agent vs. human: how do agents compare with engineers?}

We report a pilot evaluation of the first eight accepted tasks: five in operations, control and optimization, and one each in performance modeling and simulation, occupants and indoor environment, and building data. The operations tasks are RC Model Calibration, RC Model Selection, MPC Building Load Shifting, Demand-Response Test and Multi-agent RL for DER; the other three are Heat Pump Cooling Model, Day-ahead Occupancy Prediction and Weather Data Collection (Figure~\ref{fig:3}). At the time of the pilot, the platform had 18 registered users, 20 task drafts and 12 accepted tasks.

\begin{figure}[H]
  \centering
  \includegraphics[width=0.85\linewidth]{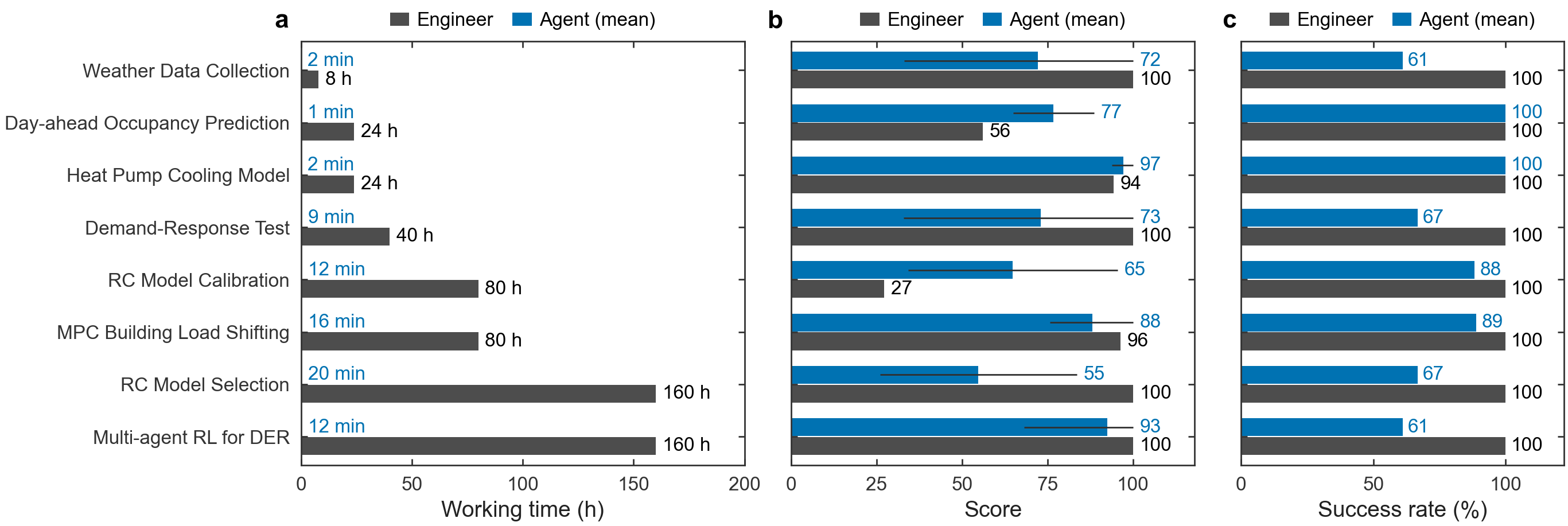}
  \caption{Agents and engineers on the eight tasks (structured request, 143 graded runs). (a) Time: the engineer's reported working time and the agents' mean run time. (b) Score: the agents' mean score and the engineer's score for each task; error bars show the standard deviation across agent runs. (c) Success rate: the share of runs that met every passing threshold.}
  \label{fig:4}
\end{figure}

\textbf{Quality.} Agents achieved scores close to those of engineers overall and scored higher on some tasks. The mean agent score was 78, compared with 84 for engineers. Agents matched or beat the engineer on three of eight tasks: RC Model Calibration, Day-ahead Occupancy Prediction and Heat Pump Cooling Model (Figure~\ref{fig:4}b). For the first two tasks, the engineers' scores were relatively low (27 and 56, respectively) because the passing thresholds were set close to their results. Overall, 79\% of agent runs succeeded, compared with a 100\% success rate for engineers (Figure~\ref{fig:4}c).

\textbf{Time.} Agents show promise for improving engineering productivity. An agent run took 9 minutes on average (median: 4 minutes), while engineers reported spending 8 to 160 working hours per task (Figure~\ref{fig:4}a).

\textbf{Cost.} The average cost per agent run was \$0.58. Across models, the average ranged from \$0.04 for GPT-5.6 Luna to \$1.31 for Claude Opus 5.5. Across tasks, the average cost per run ranged from \$0.20 to \$1.22. The full pilot included 414 runs and cost about \$253. Table~\ref{tab:3} gives the score, success rate, time and cost of each agent.

\begin{table}[htbp]
  \centering
  \small
  \caption{Results by agent on the structured request.}
  \label{tab:3}
  \begin{tabularx}{\linewidth}{X*{7}{c}}
  \toprule
  & \multicolumn{3}{c}{\textbf{Score by reasoning level}} & \multicolumn{4}{c}{\textbf{All levels}} \\
  \cmidrule(lr){2-4}\cmidrule(lr){5-8}
  \textbf{Model} & \textbf{Low} & \textbf{Medium} & \textbf{High} & \textbf{Score} & \textbf{Success} & \textbf{Time (min)} & \textbf{Cost (\$)} \\
  \midrule
  Claude Haiku 5.5 & 88 & 90 & 74 & 84 & 88\% & 14.5 & 0.23 \\
  Claude Sonnet 5.5 & 73 & 85 & 93 & 84 & 83\% & 10.6 & 0.47 \\
  Claude Opus 5.5 & 83 & 77 & 68 & 76 & 78\% & 18.4 & 1.31 \\
  GPT-5.6 Luna & 57 & 74 & 69 & 66 & 62\% & 2.5 & 0.04 \\
  GPT-5.6 Terra & 84 & 69 & 88 & 80 & 83\% & 4.6 & 0.29 \\
  GPT-5.6 Sol & 71 & 77 & 75 & 74 & 79\% & 5.4 & 1.17 \\
  \midrule
  All agents & 76 & 79 & 78 & 78 & 79\% & 9.3 & 0.58 \\
  \bottomrule
  \end{tabularx}
\end{table}

\subsection{Human to agent: does expert knowledge improve agent performance?}

\begin{figure}[htbp]
  \centering
  \includegraphics[width=\linewidth]{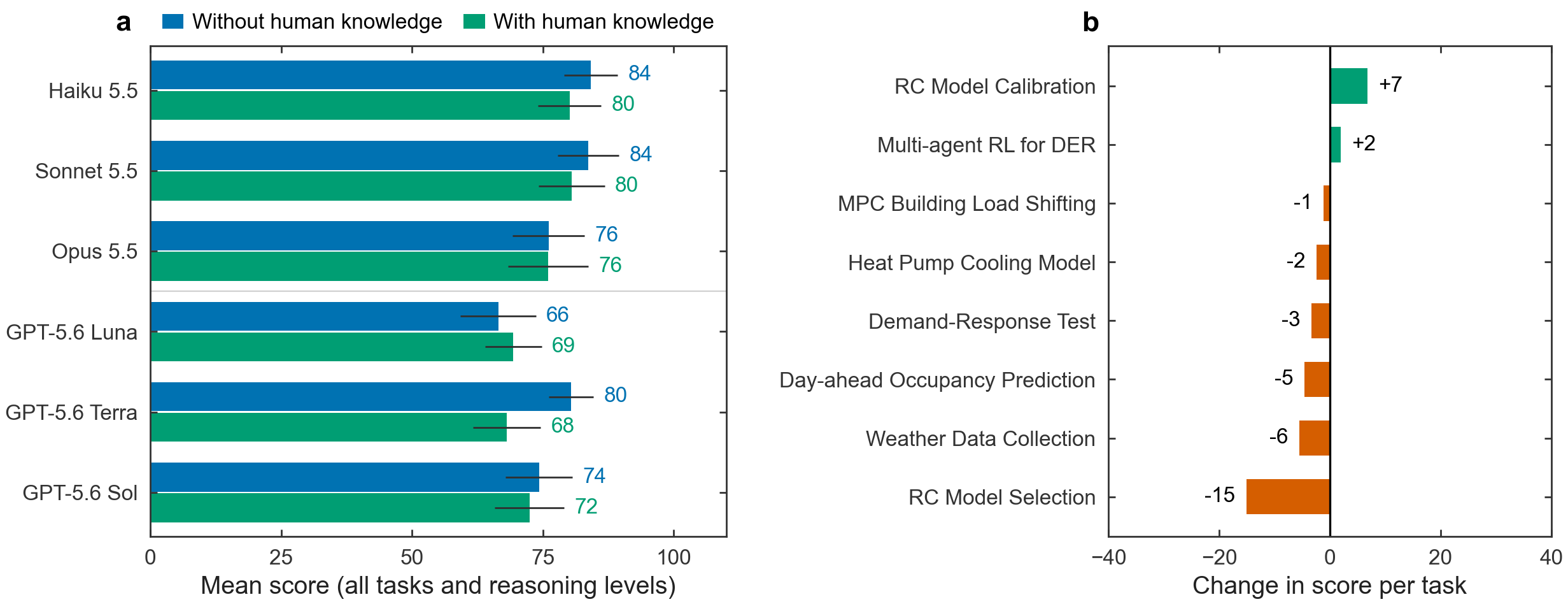}
  \caption{Effect of expert knowledge. (a) Mean score for each model without and with the contributor's knowledge, pooled across tasks and reasoning levels; error bars show the standard error. (b) Mean score change by task: the score with knowledge minus the score without it, over matched runs.}
  \label{fig:5}
\end{figure}

\textbf{Expert knowledge did not help on average.} Adding expert knowledge lowered the mean score from 77 to 74 and the success rate from 79\% to 68\%. Across 142 matched pairs, knowledge raised the score in 40 pairs, lowered it in 59, and left it unchanged in 43 (Figure~\ref{fig:5}a).

\textbf{The effect depended on the task and the agent.} Knowledge improved the mean score on two of eight tasks, most on RC Model Calibration (+7 points), and lowered it on the other six, most on RC Model Selection ($-$15; Figure~\ref{fig:5}b). Across agents, the mean change ranged from $-$12 points for GPT-5.6 Terra to +3 for GPT-5.6 Luna.

\FloatBarrier
\subsection{Human with agent: how can engineers work with agents more efficiently?}

\begin{figure}[htbp]
  \centering
  \includegraphics[width=\linewidth]{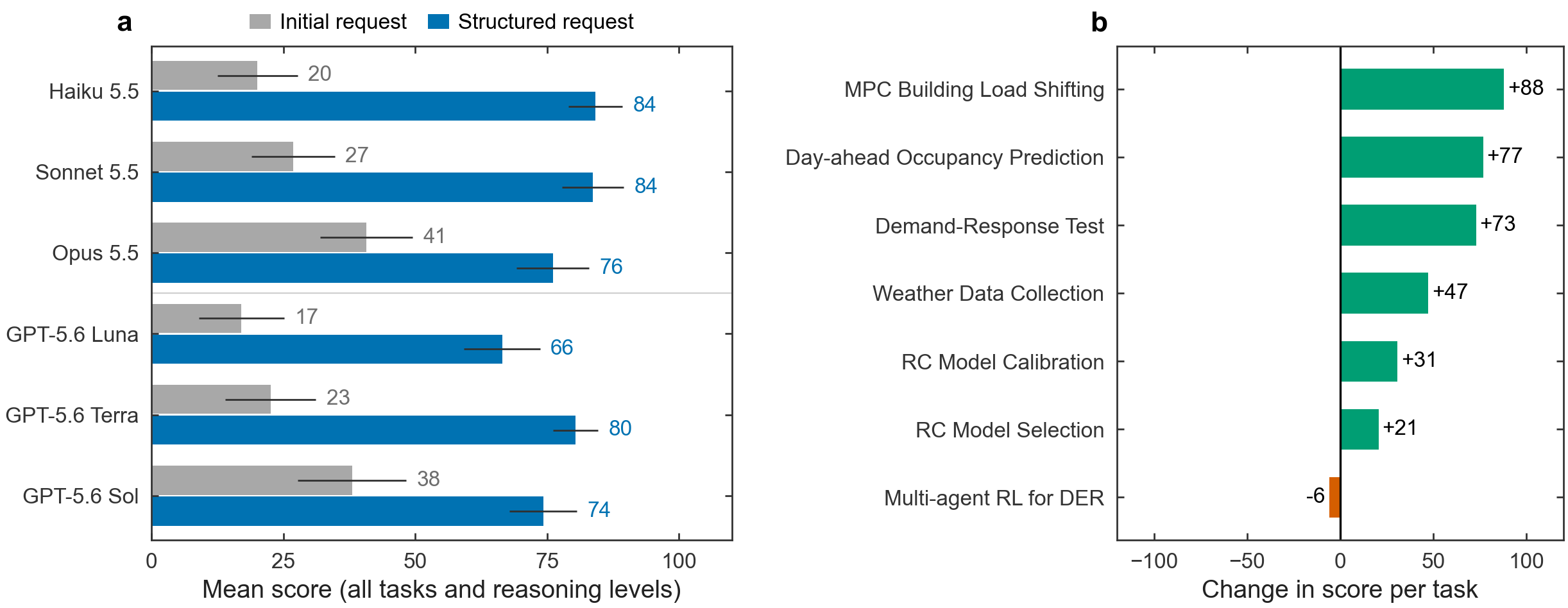}
  \caption{Effect of task specification. (a) Mean score for each model with the initial and the structured request, pooled across tasks and reasoning levels. (b) Mean score change by task: the structured-request score minus the initial-request score, over matched runs. Heat Pump Cooling Model had no separate initial request.}
  \label{fig:6}
\end{figure}

\textbf{A structured request greatly improved performance.} Over 125 matched pairs on seven tasks, the mean score rose from 27 with the initial request to 75 with the structured request, and the success rate from 5\% to 76\%. The structured request scored higher in 93 pairs, lower in 12, and the same in 20 (Figure~\ref{fig:6}). Every model improved, from +33 points for Opus 5.5 to +62 for Haiku 5.5.

\textbf{Initial requests left key requirements unclear.} Of the 120 failed runs on initial requests, the grader attributed 101 to unclear requirements. In most of them, the agent did reasonable engineering work but delivered it in a form that could not be checked: one combined file instead of seven daily files, a forecast that started at the wrong time, or missing columns. On Weather Data Collection, all nine Claude runs on the initial request retrieved the correct data but failed because the request did not ask for coordinate columns.

\begin{figure}[htbp]
  \centering
  \includegraphics[width=0.85\linewidth]{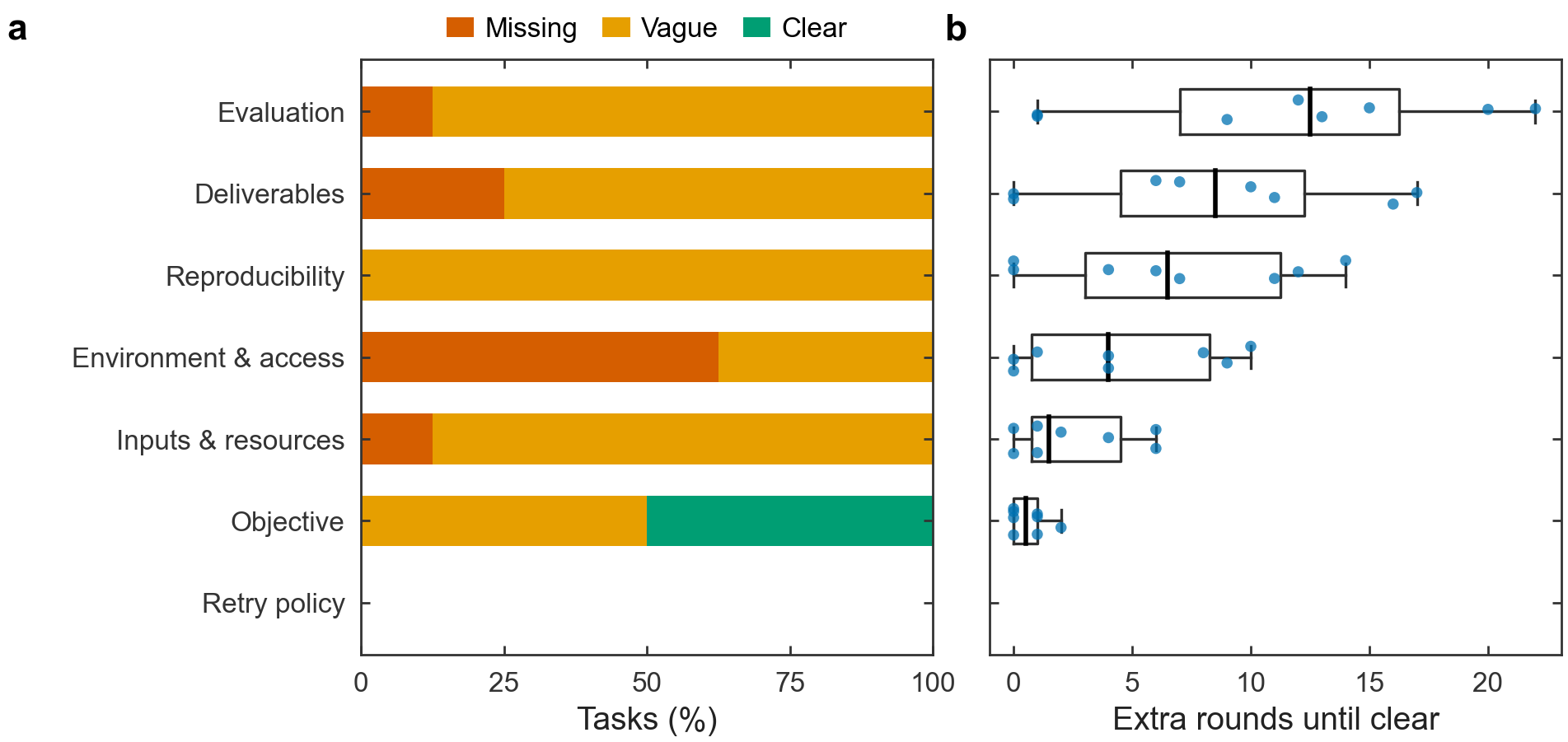}
  \caption{Completeness of initial requests. (a) Clarity of each task dimension in the contributor's initial request. (b) Additional clarification rounds needed before each dimension was rated clear; boxes show the median and interquartile range, and dots show individual tasks.}
  \label{fig:7}
\end{figure}

\begin{figure}[htbp]
  \centering
  \includegraphics[width=0.65\linewidth]{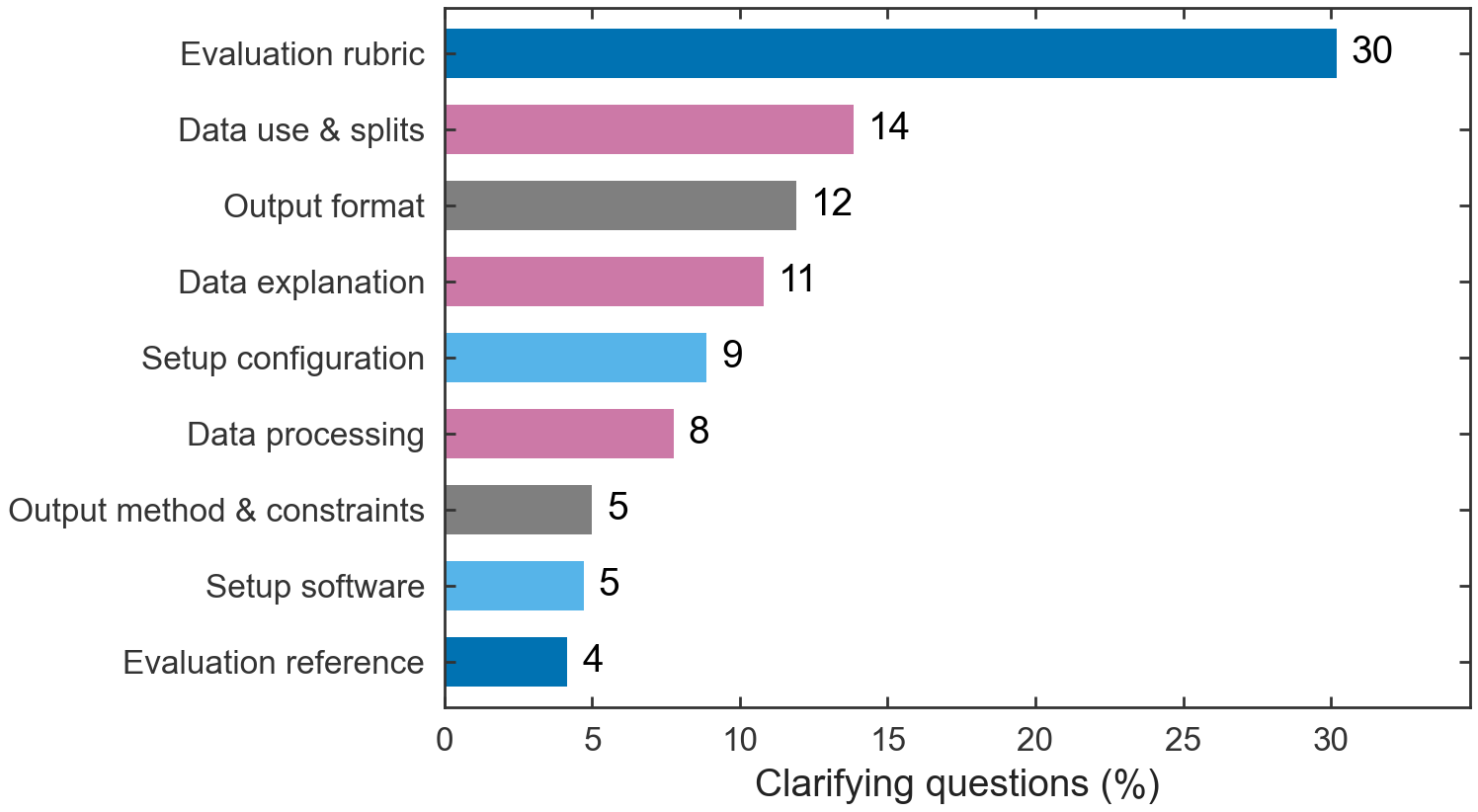}
  \caption{Categories of the 361 clarifying questions asked by the AI reviewer, as a share of all questions.}
  \label{fig:8}
\end{figure}

We analyzed the clarification records of all eight tasks. No initial request was complete. The objective was the only dimension ever rated clear, and only in four of eight tasks; every other dimension was vague or missing in all eight (Figure~\ref{fig:7}a). Evaluation needed the most clarification, with a median of 12.5 additional rounds (Figure~\ref{fig:7}b). Questions about the scoring rubric made up 30\% of the 361 clarifying questions, followed by data use and splits (14\%) and output format (12\%; Figure~\ref{fig:8}).

\textbf{Writing a clear request took substantial effort.} A complete specification took a median of 16 clarification rounds (range 4 to 24) and 65 minutes of contributor time (range 18 to 133). This was true even for experienced users: seven of the eight pilot tasks came from contributors who described themselves as agent builders. Engineers therefore benefit from a standard request format that prompts them to define objectives, inputs, outputs and evaluation criteria before agents start.

\FloatBarrier
\subsection{Task failure and lessons learned}

The pilot revealed some interesting failure behaviors where 30 failures came from the environment (11) and from the agents themselves (19; Table~\ref{tab:4}). Adding expert knowledge raised the number of failures to 45, mainly through more execution mistakes (from 9 to 15) and weaker methods (from 6 to 10). The per-task lessons explain where these failures came from.

\begin{table}[htbp]
  \centering
  \small
  \caption{Main cause of each unsuccessful run, grouped by source and diagnosed by the AI grader from the scores, outputs and transcript. \emph{Execution mistake}: a bug, wrong units or period, misread data or a wrong file format. \emph{Weak method}: a wrong or weak engineering approach. \emph{Unfinished run}: the run hit the time limit or stopped before finishing.}
  \label{tab:4}
  \begin{tabularx}{\linewidth}{X*{3}{c}}
  \toprule
  \textbf{Cause} & \textbf{Initial request} & \textbf{Structured request} & \textbf{+ Expert knowledge} \\
  \midrule
  Task: unclear requirements & 101 & 0 & 0 \\
  Environment: configuration failure & 15 & 11 & 15 \\
  Agent: execution mistake & 0 & 9 & 15 \\
  Agent: weak method & 3 & 6 & 10 \\
  Agent: unfinished run & 1 & 4 & 5 \\
  \midrule
  Unsuccessful runs & 120 & 30 & 45 \\
  Graded runs & 126 & 143 & 142 \\
  \bottomrule
  \end{tabularx}
\end{table}

\textbf{Tool access strongly affects task success.} On Weather Data Collection, the GPT sandbox had networking disabled, making the weather API unreachable. On Demand-Response Test, GPT agents replaced EnergyPlus simulations with surrogate models. Claude agents searched the machine, found an installed EnergyPlus build, and scored 100 on every run with a structured request. The affected runs need to be repeated after the environment is correctly configured.

\textbf{Stronger models make failure less visible.} When a required simulator or API was unavailable, several agents still produced placeholder results in the required format. On Weather Data Collection, nine runs filled the files with placeholder values that passed format checks and scored 50. On Demand-Response Test, eight structured-request runs delivered surrogate results and a placeholder FMU. The files looked valid even though the required data collection or simulation had not been completed. This raises an important question about how engineers should verify agent outputs. As agents become more capable, we may trust their results more readily and spend less time checking them. This issue deserves further investigation as more cases become available.

\textbf{Expert knowledge narrowed the agents' search.} Some agents followed the provided guidance too closely. On Heat Pump Cooling Model, 15 of 18 runs given the contributor's workflow followed it step by step and reproduced the reference score of 94.4. Eleven runs that examined residuals at each outdoor temperature identified a low-ambient operating mode missed by the contributor's method and scored 99.5 to 100, outperforming the engineer. On RC Model Selection, the added guidance encouraged heavier cross-validation and three-node models, and success fell from 12 to 6 of 18 runs.

\section{Discussion}
\label{sec:discussion}

\textbf{Task specification is the main bottleneck.} The largest improvement in the pilot came from how tasks were specified, rather than differences in model capability. With initial requests, agents often performed reasonable engineering work but produced outputs that did not meet the expected deliverable or evaluation requirements. Structured requests eliminated this type of failure (Table~\ref{tab:4}). This suggests that effective delegation requires more effort in task specification. In the pilot, contributors spent a median of 65 minutes preparing each structured request, which is relatively small compared with the 8 to 160 hours required by engineers to complete the tasks.

\textbf{The effect of expert knowledge is not as expected.} The pilot suggests that expert knowledge can improve performance when it defines requirements and validation checks, but may reduce performance when it prescribes specific methods. Agents often followed the provided methods even when better alternatives were available. This highlights the need to evaluate expert knowledge, including knowledge packaged as agent skills~\cite{anthropic2025skills}, through paired runs before deployment. However, these findings remain preliminary.

\textbf{Efficient human-agent collaboration needs more structured task deliverable formats.} The findings highlight the importance of clear task specifications and reliable validation when working with AI agents. Engineers should clearly define task inputs, expected deliverables, evaluation criteria, and validation requirements. Expert knowledge can then be provided as guidance on what to achieve and verify, while allowing agents to explore different methods. Preparing the necessary tools and execution environments can also reduce failures caused by missing dependencies. Finally, agent outputs should be evaluated against the specified criteria before being used in engineering applications.

\section{Conclusion}
\label{sec:conclusion}

We introduced Buildrix, an open benchmark that evaluates AI agents on real building engineering tasks and compares them with the engineers who did the work. In a pilot of eight tasks and 414 runs, agents scored close to engineers in a small fraction of the time and at low cost. The largest effect came not from the model or the reasoning level but from the task specification: a clear, structured request raised the mean score from 27 to 75. Expert knowledge did not improve the average and sometimes held agents to the level of the method it described. For current agents, the main barrier to taking on building engineering work appears to be how the work is specified, not their capability. As Buildrix grows, we will test these findings across all eight domains and turn the reviewed lessons into an open knowledge base and reusable agent skills for the building community.

\end{document}